\documentclass[a4paper]{jpconf}
\usepackage{graphicx}
\usepackage{amssymb}
\begin{document}
\title{Measurements of $t$-channel single top-quark production cross sections at $\sqrt{s} = 7$ TeV with the ATLAS detector}

\author{Dominic Hirschb\"uhl \\ On behalf of the ATLAS Collaboration}

\address{Gausstr. 20, 42119 Wuppertal, Germany}

\ead{hirsch@physik.uni-wuppertal.de}

\begin{abstract}
This article presents measurements of the t-channel single top-quark $(t)$ and top-antiquark $(\bar{t})$ total production
cross sections $\sigma(tq)$ and $\sigma(\bar{t}q)$, their ratio $R_t=\sigma(tq)/\sigma(\bar{t}q)$.
Differential cross sections for the  $\sigma(tq)$ and $\sigma(\bar{t}q)$ processes are measured
as a function of the transverse momentum and the absolute value of the rapidity of  $(t)$ and $(\bar{t})$, respectively. 
The analysed data set was recorded with the ATLAS detector and corresponds to an integrated luminosity of 4.59 fb$^{−1}$.
The cross sections are measured by performing a binned maximum-likelihood fit to the output distributions of neural networks. 
The resulting measurements are  $\sigma(tq)=46\pm 6$ pb, $\sigma(\bar{t}q)=23 \pm 4$ pb, $R_t=2.04 \pm 0.18$, 
consistent with the Standard Model expectation.
\end{abstract}

\section{Introduction}
In proton--proton ($pp$) collisions at the LHC, top quarks are produced at unprecedented rates.
Single top-quark production is described by three subprocesses that are distinguished by the virtuality of the exchanged $W$ boson. 
The dominant process is the $t$-channel exchange.
A light quark from one of the colliding protons interacts with a $b$-quark from 
another proton by exchanging a virtual $W$ boson ($W^*$). 
Since the $u$-quark density of the proton 
is about twice as high as the $d$-quark density,
the production cross section of single top quarks $\sigma(tq)$ is expected to be about twice as high
as the cross section of top-antiquark production $\sigma(\bar{t}q)$.
In $pp$ collisions at $\sqrt{s}=7$ TeV,
the total inclusive cross sections of top-quark and top-antiquark production in the $t$-channel  are predicted to be
$\sigma(tq)  =  41.9 ^{+1.8}_{-0.9}\ \mathrm{pb}, \; \sigma(\bar{t}q) = 22.7 ^{+0.9}_{-1.0}\ \mathrm{pb}$
with approximate next-to-next-to-leading-order (NNLO) precision~\cite{Kidonakis:2011wy}.


\section{Event selection and background estimation}
\begin{figure}[h]
\vspace{-9pc}
\includegraphics[width=16pc]{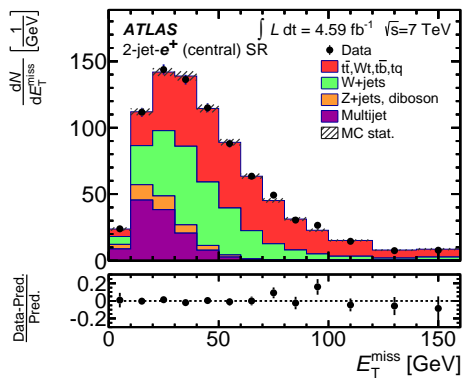}\hspace{2pc}%
\begin{minipage}[b]{18pc}\caption{\label{fig:missetfit} $E_{\mathrm{T}}^{\mathrm{miss}}$~distributions in the signal region~(SR) 
     for the 2-jet-$e^+$  channel for central electrons. The distributions are normalized to the result of a binned maximum-likelihood fit.  
     The relative difference between the observed and expected number of events 
     in each bin is shown in the lower panels.}\vspace{3pc}%
\end{minipage}
\end{figure}

The analysis described in this article~\cite{Aad:2014fwa} uses $pp$ LHC collision data collected at a
center-of-mass energy of 7 TeV with the ATLAS detector~\cite{ATL-2008-001_sgtop}.
Stringent detector and data quality requirements are applied, resulting in a data set 
corresponding to an integrated luminosity of $4.59\pm0.08$~fb$^{-1}$.
The event selection requires exactly one charged isolated lepton ($e$ or $\mu$), 
exactly two or three jets, and $E_{\mathrm{T}}^{\mathrm{miss}}>30$ GeV.
Electron candidates are selected from energy deposits  in the LAr electromagnetic calorimeter matched to 
tracks and are required to have $E_{\mathrm{T}} >25$ GeV and $|\eta|<2.47$.
Muon candidates are reconstructed by combining track segments found in the inner detector and the muon spectrometer.
Only candidates that have $p_\mathrm{T}>25$ GeV and $|\eta|<2.5$ are considered.
Jets are reconstructed using the anti-$k_{t}$ algorithm with a radius parameter of 0.4 and have to have 
$p_\mathrm{T} >30$ GeV and $|\eta|<4.5$. 
Jets in the endcap/forward-calorimeter transition region, corresponding to 
$2.75<|\eta|<3.5$, must have $p_\mathrm{T} >35$ GeV.
At least one of the jets must be $b$-tagged.
Since the multijet background is difficult to model precisely, its contribution is reduced by 
$m_\mathrm{T}\left(\ell E_{\mathrm{T}}^{\mathrm{miss}} \right)=\sqrt{2 p_\mathrm{T}(\ell) \cdot E_{\mathrm{T}}^{\mathrm{miss}}
  \left[1-\cos \left( \Delta \phi \left(\ell, E_{\mathrm{T}}^{\mathrm{miss}} \right) \right) \right]} > $ 30 GeV and
by requiring
$\Delta \phi \left(j_1, \ell \right)$ $p_\mathrm{T} \left(\ell\right) > 40\, {\mathrm GeV} \cdot \left(1 -  \frac{\pi - |\Delta \phi\left(j_1, \ell \right)|}{\pi -1} \right)$, where $j_1$ denotes the leading jet.
The $W$+jets background is initially normalized to the theoretical prediction and then subsequently
determined simultaneously both in the context of the multijet background estimation and as part of the extraction 
of the signal cross section. The estimated number of events of the theoretically well know ($t\bar{t}$, $Wt$, $s$-channel
single top-quark production) or small processes ( $WW$, $WZ$ and $ZZ$, $Z$+jets background) are calculated using the theoretical prediction.
Data driven techniques are used to derive the multijet background.
In the electron channel the jet-lepton method is used. There an electron-like jet is selected with special requirements 
and redefined as a lepton.
A binned maximum-likelihood fit to observed data in the $E_{\mathrm{T}}^{\mathrm{miss}}$~distribution, omitting the  $E_{\mathrm{T}}^{\mathrm{miss}}$~requirement in the selection, to obtain the normalization, as shown in \Fref{fig:missetfit}.
\\
In the muon channel, the matrix method is used to obtain both the normalization and shape of the multijet background.
The method estimates the number of multijet background events in the signal region based on loose and tight lepton isolation
definitions, the latter selection being a subset of the former.

\section{Signal extraction}
\begin{figure}[h]
\vspace{-11pc}
\begin{minipage}[b]{18pc}
     \includegraphics[width=18pc]{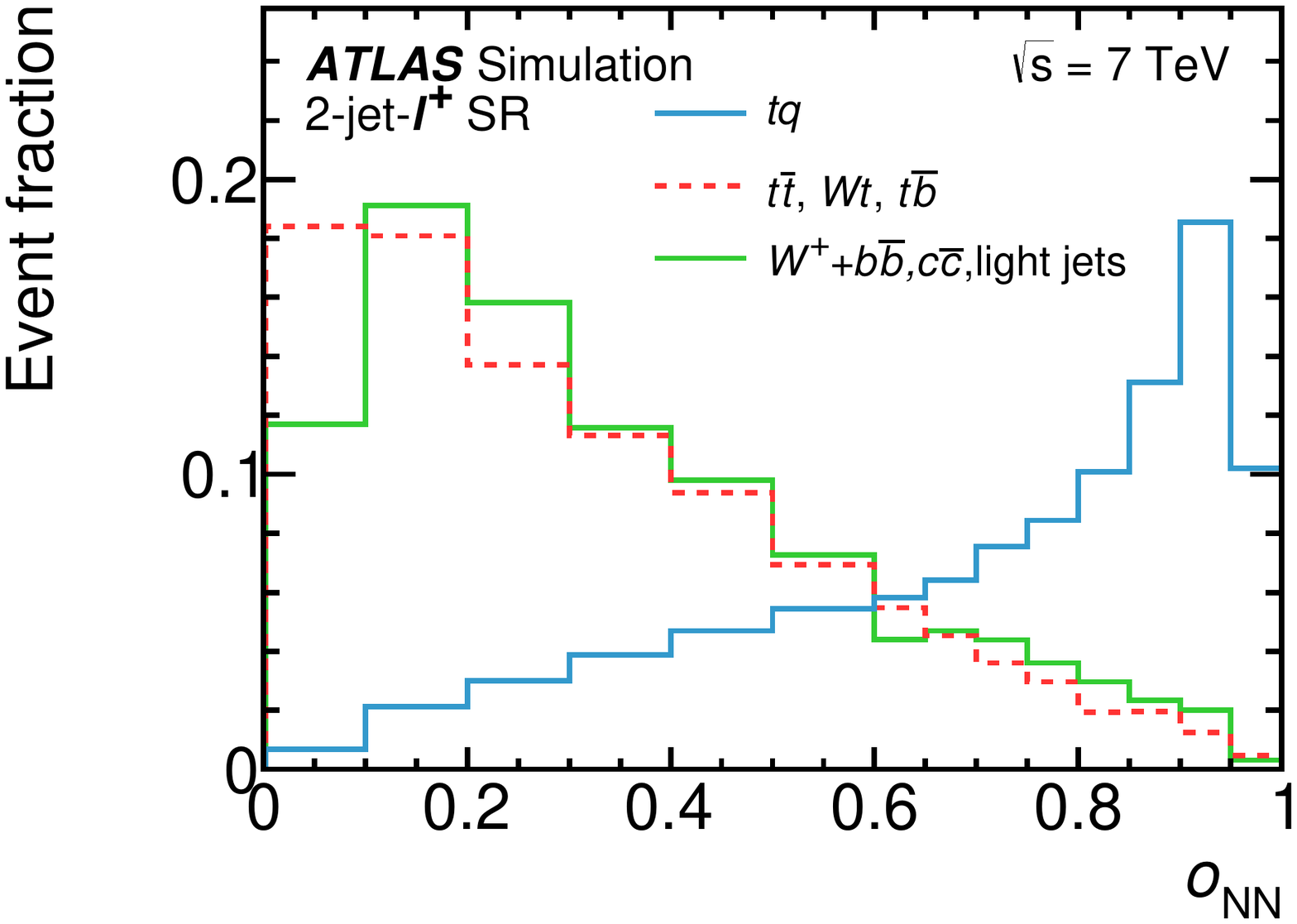}
     \caption{\label{fig:nn_templates} Probability densities of the NN discriminants
     in the 2-jet-$\ell^+$ channel in the signal region (SR). 
     The distributions are normalized to unit area.}\vspace{3pc}%
\end{minipage}\hspace{2pc}%
\begin{minipage}[b]{18pc}
  \includegraphics[width=18pc]{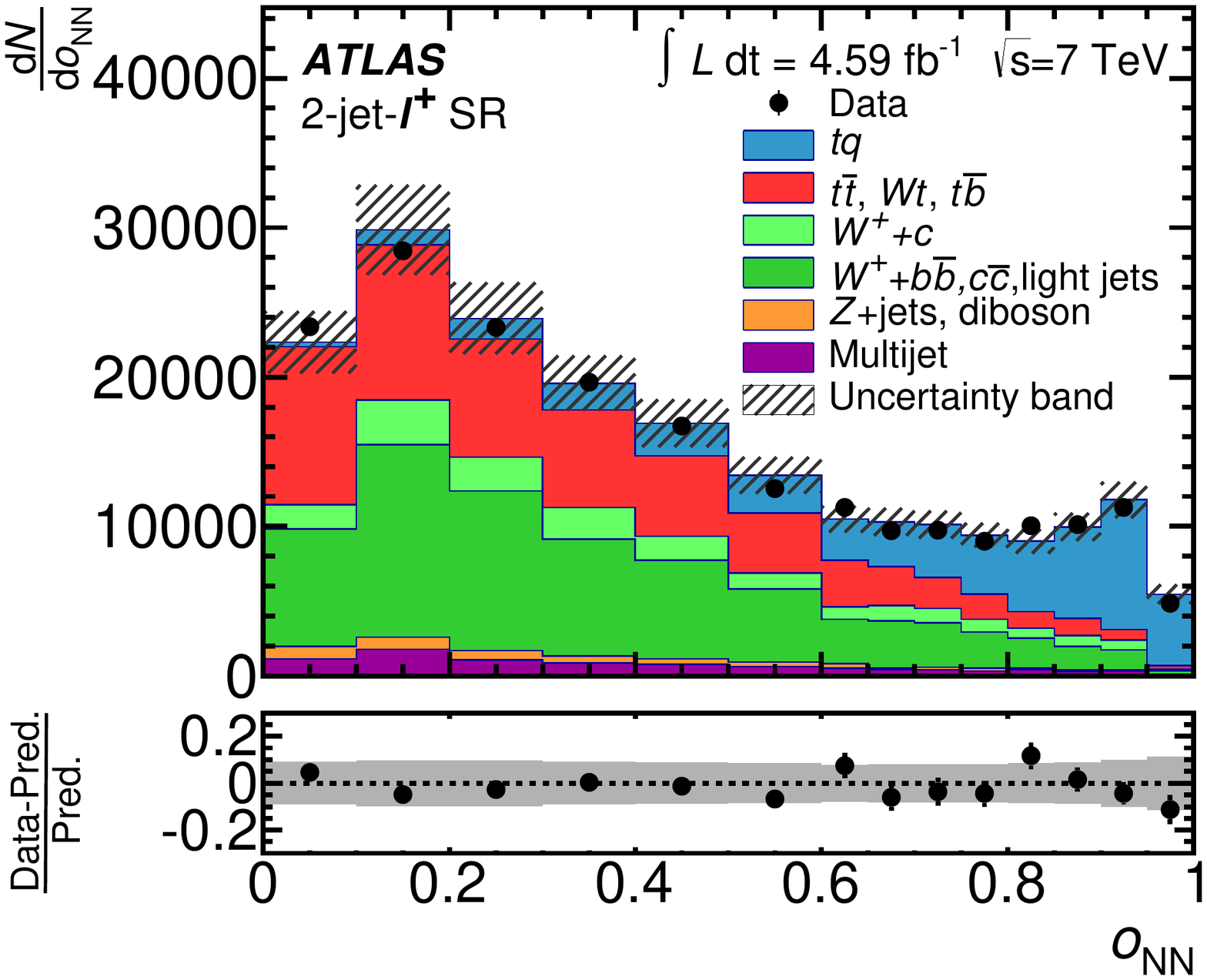}
  \caption{\label{fig:nnout_fit_results} 
  Neural network discriminant distributions normalized to the fit result in the 2-jet-$\ell^+$ channel.
  The relative difference between the observed and expected number of events is shown in the lower panels.}\vspace{2pc}%
\end{minipage}
\end{figure}

To separate $t$-channel single top-quark signal events from 
background events, several kinematic variables are combined to form powerful discriminants by
employing neural networks. The NeuroBayes~\cite{Feindt:2006pm} tool is used
for preprocessing the input variables and for the training of the NNs.
A large number of potential input variables were 
studied and the correct modeling of the variables is checked in a control region.
In \Fref{fig:nn_templates}, the probability densities of the resulting NN 
discriminants in one channel for the signal and the most important backgrounds is shown.

The cross sections $\sigma(tq)$ and $\sigma(\bar{t}q)$ are extracted by performing a binned 
maximum-likelihood fit to the NN discriminant distributions in the 
2-jet-$\ell^+$, 2-jet-$\ell^-$, 3-jet-$\ell^+$-1-tag, and 3-jet-$\ell^-$-1-tag channels and 
to the event yield in the 3-jet-2-tag channel, treating 
$t$-channel top-quark and $t$-channel top-antiquark production as independent processes.
The cross-section ratio is subsequently computed as $R_t = \sigma(tq)/\sigma(\bar{t}q)$. 
In \Fref{fig:nnout_fit_results} the observed NN discriminant distribution in the 2-jet-$\ell^+$ channel
is shown compared to the compound model of signal and background normalized to the fit results. 

\section{Systematic uncertainties}
Systematic uncertainties are assigned to account for detector calibration and resolution uncertainties, as well as the  
uncertainties of theoretical predictions.
Uncertainties on the reconstruction and energy calibration of 
jets, electrons and muons are propagated through the entire analysis.
Systematic uncertainties arising from the modeling of the single top-quark signal,
the $t\bar{t}$ background, and the $W$+jets background are taken into account as well as 
uncertainties related to the PDFs.\\
The systematic uncertainties on the individual top-quark and top-antiquark 
cross-section measurements and their ratio are determined using pseudo-experiments
that account for variations of the signal acceptance, the background rates, and the shape of the NN discriminant. 
The correlations between the different channels and the physics processes are fully accounted for.
The dominant systematic uncertainty on the cross sections is the JES $\eta$-intercalibration
uncertainty.

\section{Total cross-section measurements}
\begin{figure}[h]
\vspace{-14pc}
\includegraphics[width=18pc]{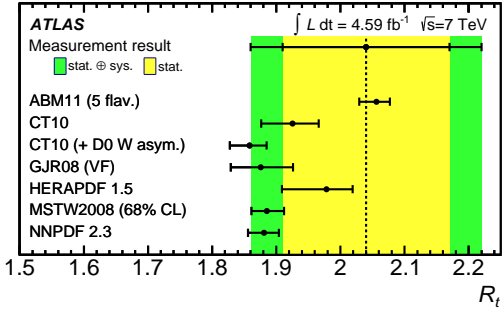}\hspace{2pc}%
\begin{minipage}[b]{18pc}\caption{\label{fig:rtop} Comparison between observed and predicted values of  $R_t$. 
  The predictions are calculated at NLO precision in the five-flavor scheme and 
  given for different NLO PDF sets and the uncertainty includes the uncertainty on the renormalization and factorization scales, 
  the combined internal PDF and $\alpha_{\mathrm{s}}$ uncertainty. }\vspace{1pc}%
\end{minipage}
\end{figure}

After performing the binned maximum-likelihood fit and estimating the total uncertainty the results are:
$$\setlength\arraycolsep{0.1em}
 \begin{array}{rclcl}
  \sigma(tq)       & = & 46 \pm 1\, (\mathrm{stat.}) \pm 6\, (\mathrm{syst.})\, \mathrm{pb} & = & 46\pm 6\, \mathrm{pb}, \\
  \sigma(\bar{t}q) & = & 23 \pm 1\, (\mathrm{stat.}) \pm 3\, (\mathrm{syst.})\, \mathrm{pb} & = & 23 \pm 4\, \mathrm{pb} \ \ \ \mathrm{and}  \\
  R_t             & = & 2.04 \pm 0.13\, (\mathrm{stat.})\, \pm 0.12\, (\mathrm{syst.}) & = & 2.04 \pm 0.18.
 \end{array}
$$
\Fref{fig:rtop} compares the measured values of $R_t$ to NLO predictions from
MCFM~\cite{Campbell:2009ss} and Hathor~\cite{Kant:2014oha} using different PDF sets. 
\\
\vspace{-2pc}
\section{Differential cross-section measurements}
\begin{figure}[h]
\vspace{-7pc}
\begin{minipage}{18pc}
\includegraphics[width=18pc]{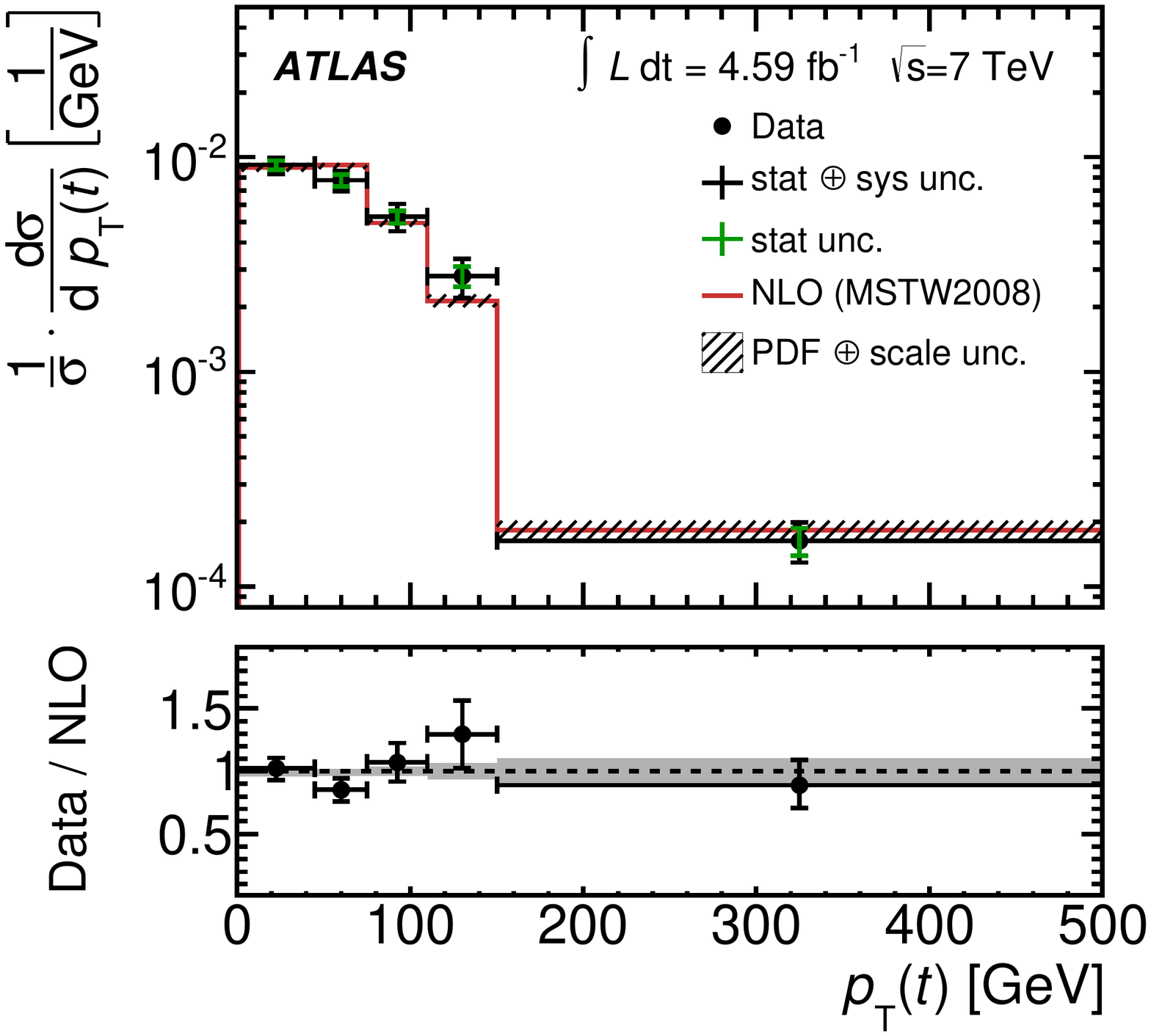}
\caption{\label{fig:toppt} Differential cross section as a function of $p_\mathrm{T}(t)$. }
\end{minipage}\hspace{2pc}%
\begin{minipage}{18pc}
\includegraphics[width=18pc]{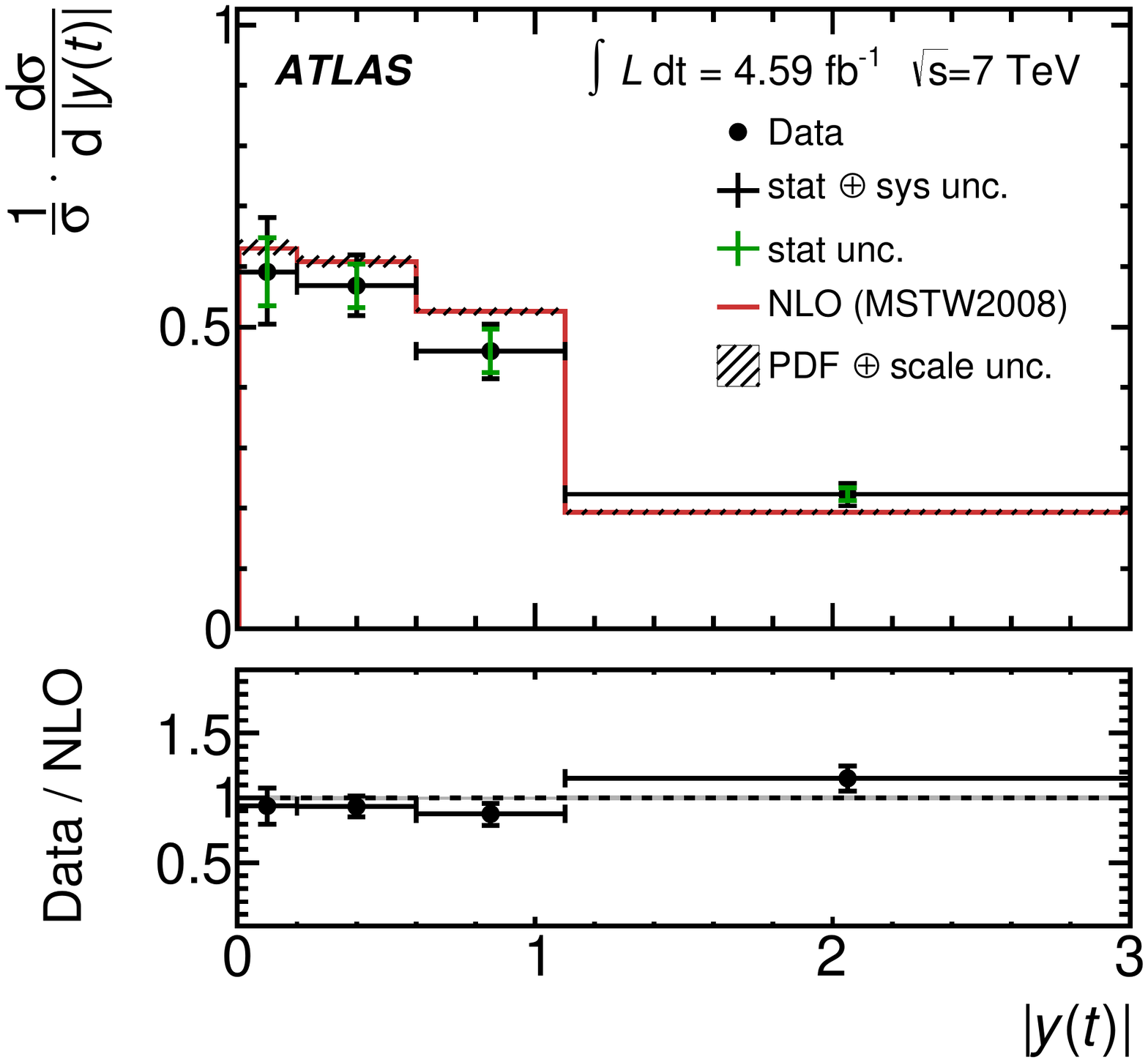}
\caption{\label{fig:toprap}Differential cross section as a function of $y(t)$. }
\end{minipage} 
\end{figure}

A high-purity region (HPR) is defined to measure the differential cross sections in the 2-jet-$\ell^+$ 
and 2-jet-$\ell^-$ channels, by requiring the NN discriminant to be larger than~$0.8$. 
In the 2-jet-$\ell^+$ HPR the signal contribution is twice as large as the background contribution, while
it is approximately the same size in the 2-jet-$\ell^-$ HPR. 
Differential cross sections are measured as a function of the $p_\mathrm{T}$ and $|y|$ of $t$ and $\bar{t}$. 
The binning of the differential cross sections is chosen based on the experimental resolution of the $p_\mathrm{T}$ and $|y|$ distributions
as well as the data statistical uncertainty. 
The measured distributions are distorted by detector effects and acceptance effects therefore
the observed distributions are unfolded to the four-momenta of the top quarks before the decay 
and after QCD radiation.
A migration matrix is built by relating the variables at the reconstruction and at the parton level using the signal simulation
and  the selection efficiency of each variable is defined  as the ratio of the parton-level yield before and after selection. 
A graphical representation of the unfolded top quark $p_\mathrm{T}$ distribution is shown in \Fref{fig:toppt} 
and for the rapidity in \Fref{fig:toprap}.  Both distributions are compared to NLO predictions from 
MCFM using the MSTW2008 PDF set.
For of the normalized differential cross sections many systematic uncertainties cancel and thus the measurement is dominated by the
statistical uncertainty and the uncertainty due to the Monte Carlo sample size.

\section*{References}
\bibliographystyle{iopart-num}
\bibliography{mybib}
\end{document}